# Year after year: Tortured conference series thriving in Computer Science

Wendeline Swart[1], Guillaume Cabanac[1,2]

[1] Université Toulouse 3 – Paul Sabatier, IRIT UMR 5505 CNRS, Toulouse, France
[2] Institut universitaire de France (IUF), Paris, France

**Objective.** The 'Problematic Paper Screener' (PPS, WCRI'22, https://doi.org/10.48550/arXiv.2210.04895) flagged 12k+ questionable articles featuring tortured phrases, such as 'glucose bigotry' instead of 'glucose intolerance.' It daily screens the literature for 'fingerprints' from a list of 4k tortured phrases known to reflect nonsensical paraphrasing with synonyms. We identified a concentration of 'tortured articles' in IEEE conferences and reported our concerns in November 2022 (https://retractionwatch.com/?p=127299). This WCRI submission unveils 'tortured conference series': questionable articles that keep being accepted in successive conference editions.

**Method.** We analysed data from the 'Tortured detector' of the PPS (https://bit.ly/PPS-tortured). The considered corpus includes articles from conference proceedings flagged with 5+ tortured phrases ($n$ = 3848). We grouped the records by publisher and by conference. For each conference, we calculated the yearly (i.e., per edition) number of tortured articles included in the conference proceedings.

**Results.** The 3848 tortured articles were published over 2008–2023 in the conference proceedings of 11 publishers. Those publishing 10+ tortured articles are: IEEE ($n$ = 3227; 84%), IOP Publishing ($n$ = 314; 8%), AIP Publishing ($n$ = 233; 6%), ACM ($n$ = 45; 1%), and Atlantis Press ($n$ = 10; <1%). IEEE (Institute of Electrical and Electronics Engineers) and ACM (Association for Computing Machinery) are the two flagship professional associations of computer scientists: they award the two major prizes in Computer Science (CS) and publish most of CS research. The 45 tortured ACM articles appeared in 35 conference editions, with a top concentration of 4 articles in AICTC'16. In contrast, way higher concentrations affected IEEE: up to 17% ($n$ = 44) tortured articles appeared in the ICERECT'22 proceedings. Even more troubling: we flagged 172 conference series with 2+ tortured articles published over 2+ editions. The worst cases in volume and longevity are 1) ICACCS with 82 tortured articles over 6 years since 2017 and 2) ICCES with 75 tortured articles over 8 years since 2016.

**Conclusion.** IEEE has published the highest number of conference proceedings featuring tortured articles. We found evidence of affected conference series and advise IEEE to audit their peer review process to prevent further contamination of the literature.

**Acknowledgements.** GC acknowledges the NanoBubbles project that has received Synergy grant funding from the European Research Council (ERC), within the European Union's Horizon 2020 program, grant agreement no. 951393.